\documentclass{aa}
\usepackage{psfig}

\usepackage{graphics}
\usepackage{psfig}
\begin{document}
\thesaurus{11(08.01.1,08.09.2,08.09.2,11.01.1,11.09.1,11.04.2)}
\renewcommand{\dbltopfraction}{.99}
\renewcommand{\dbltextfloatsep}{0.001cm}
\setcounter{dbltopnumber}{10}

\title{First results of UVES at VLT: abundances 
in the Sgr dSph
\thanks{Based on 
public data released from the UVES commissioning at the VLT Kueyen 
telescope, European Southern Observatory, Paranal, Chile.
}
}
%\subtitle{}
\author{Piercarlo Bonifacio\inst{1}
\and Vanessa Hill\inst{2}
\and Paolo Molaro\inst{1}
\and Luca Pasquini\inst{2}
\and Paolo  Di Marcantonio\inst{1}
\and Paolo Santin\inst{1}
}
\offprints{P. Bonifacio}
\institute{
Osservatorio Astronomico di Trieste,Via G.B.Tiepolo 11, 
I-34131 Trieste, Italy
\and
European Southern Observatory
K. Schwarzschild Strasse 2 D-85748 Garching bei M\"unchen - Germany}
\mail{bonifaci@ts.astro.it}
\date{received .../Accepted...}
\maketitle

\begin{abstract}
Two giants 
of the Sagittarius dwarf spheroidal
have been observed 
with the UVES spectrograph on the ESO 8.2m Kueyen telescope, during
the commissioning of the instrument.
Sgr 139 has  [Fe/H]$=-0.28$
and Sgr 143 [Fe/H]$=-0.21$, these values are considerably higher
than photometric estimates of the metallicity of the main population of Sgr.
We derived abundances for O, Na, Mg, Al, Si,
Ca, Sc, Ti, V, Cr, Mn, Co, Ni, Cu, Y, Ba, La, Ce, Nd and Eu;
the abundance ratios found are 
essentially solar with a few  exceptions: Na shows a strong
overdeficiency, the heavy elements Ba to Eu, are
overabundant, while Y is underabundat.
The high metallicity derived implies that the 
Sgr galaxy has experienced a high level of chemical
processing.
The stars had been selected to be representative
of the two main stellar populations of Sagittarius, however,
contrary to what expected from the photometry, the two stars
show a very similar  chemical composition.
We 
argue that the most likely
explanation for the difference in the photometry of the two stars 
is a different distance, Sgr 143 being about 2Kpc nearer 
than Sgr 139.
This result suggests that the interpretation of  
colour -- magnitude diagrams of Sgr is more complex
than previously thought and  the effect of
the line of sight depth should not be neglected.
It also shows that spectroscopic abundances are required for
a correct interpretation of Sgr populations.
\end{abstract}

\begin{figure*}
\psfig{figure=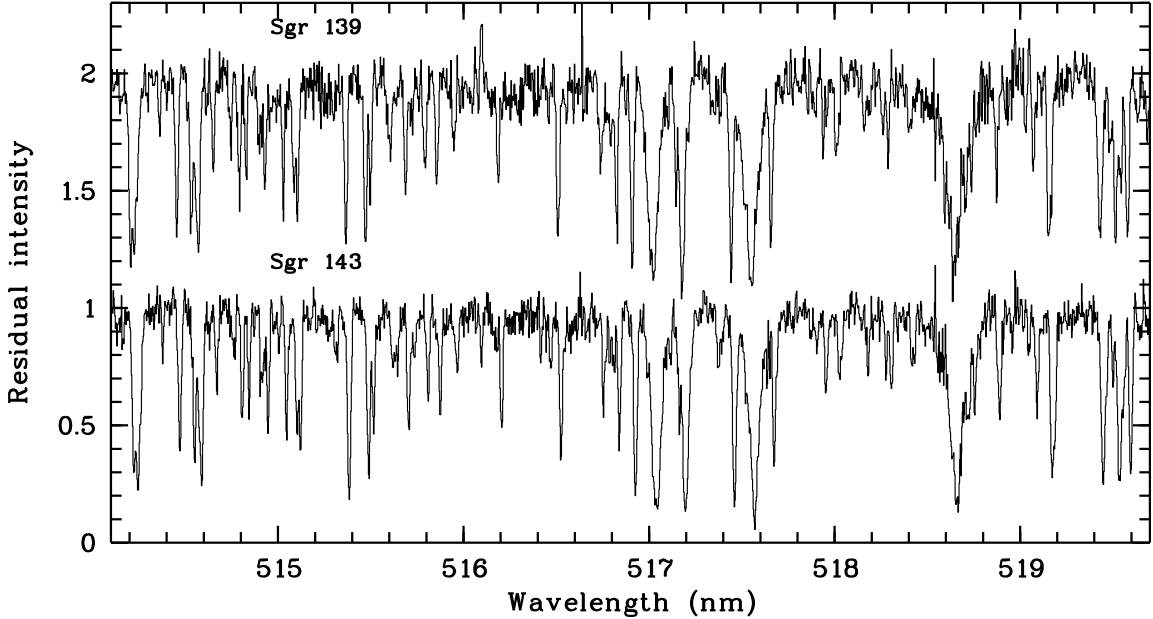,width=15.2cm,clip=t}
\caption{The spectra of the two stars
in the region of the Mg I b triplet. The spectrum
of Sgr 139 has been shifted vertically by one unit for display
purposes. Wavelengths are observed wavelengths, not shifted to
rest.}
\label{spectra}
\end{figure*}

\section{Introduction}

In the recent years our ideas on galaxy formation and evolution 
have considerably developed,
and it is generally acknowledged that 
it is a complex process which may well take different paths in different
galaxies. Much attention is being devoted to dwarf spheroidal
galaxies, essentially for two reasons: 1) they seem to be relatively
simple systems, typically characterized by a single stellar population;
in such a system we hope to be able to isolate some of the key
ingredients of the phenomenon; 2) Interaction of these dwarf
galaxies with large galaxies (such as our own or the Andromeda galaxy)
could, in principle, play an important role in shaping the
morphology of the large galaxies. The nearest
members of the class, the dwarf spheroidals of the Local Group, are
close enough that their stars are  amenable to  detailed  analysis
with the same techniques employed to study Galactic stars, with the
advent of the new 8m class telescopes.  
In this paper we report on such an observation: the first
detailed chemical 
analysis of two stars in the Sgr dwarf spheroidal 
based on high resolution spectra obtained with the UVES spectrograph
on the ESO  8.2m Kueyen telescope.
%The spectra were secured during the first commissioning of UVES.
Ever since the discovery
of Sgr (Ibata et al 1994) 
photometric studies have shown the red giant branch (RGB)
of Sgr to be wider than expected for a population with a single
age and metallicity. 
This has been generally
interpreted as evidence  that Sgr displays a spread in metallicity
which is likely due to different bursts of star formation.
Ibata et al (1995) found a mean metallicity
of [Fe/H]=-1.25 and their metallicity distribution
displays a spread of over 1 dex.
Sarajedini \& Layden (1995)
found a main population with [Fe/H]= -0.52 
and suggested the possible
existence of a population of [Fe/H]$\approx -1.3$.
Mateo et al (1995) provide a mean metallicity of $-1.1 \pm 0.3$,
Ibata et al (1997) estimate
metallicities in the range $-1.0 -0.8 $, Marconi et al (1998)
$ -1.58 -0.7 $ , Bellazzini et al (1999) $-2.1, -0.7$. 
The age of Sgr may not be disentangled from its metallicity,
from Main Sequence fitting, Fahlman et al (1996) found
acceptable solutions for an age of 10 Gyr and metallicity $-0.8$
or an age of 14 Gyr and a metallicity $-1.3$.
Clearly if Sgr may not be described as a single population
the concepts of age and metallicity loose some
of their significance; Bellazzini et al (1999) proposed an
extreme scenario in which star formation began rather
early and continued for a period longer than
4 Gyr. 
Foreseeing the potentiality of UVES to perform detailed abundance analysis
of these stars, to confirm or refute 
the photometrically inferred spread in 
metallicity, we undertook already in 1995 observations of 
low resolution spectra of photometrically identified (Marconi et al 1998)
Sgr candidates. The main purpose was to obtain confirmed radial
velocity members of Sgr for subsequent high resolution follow-up
with UVES. From the low resolution spectra we also devised a method
to obtain crude metallicity estimates from spectral indices
defined in the Mg I b triplet region.
The two stars were selected from this low-resolution study
of Sgr
with photometry and abundance estimates
which suggested these stars to differ by at least   
0.5 dex in metallicity.

\begin{table}[b]
\caption{Basic Data}
\begin{center}
\renewcommand{\tabcolsep}{0.05cm}
\begin{tabular}{lrrrrrrr}
\hline
\# & $\alpha_{2000}$ & $\delta_{2000} $ & $V$ \hfill& $(V-I)_0$ &$T_{eff}$
& log g  & $\xi$\\
&  & & & & K           &         & kms$^{-1}$\\
\hline 
139 & 18 53 50 & $-30$ 30 45 & 18.33 & 0.965 &4902&2.5 &1.4\\
143 & 18 53 49 & $-30$ 31 60 & 18.15 & 0.947 &4932&2.5 & 1.5\\
\hline 
\end{tabular}
\end{center}
\end{table}

\begin{table}[t]
\caption{Mean abundances}
\begin{center}
\begin{tabular}{lrrrrrr}
\hline
& [X/H] & $\sigma$ & n  &[X/H] & $\sigma$ & n \\ 
& 139   &          &    & 143  &          &   \\
\hline
O I   & $<-0.53 $ & --    & 1 & $-0.37$  &      & 1 \\
Na I  & $ -0.61 $ & 0.12  & 2 & $ -0.78 $ & 0.17 & 3 \\ 
Mg I  & $ -0.33 $ & 0.11  & 2 & $ -0.44 $ & 0.12 & 3 \\
Al I  & $ -0.52 $ & 0.08  & 3 & $ -0.40 $ & 0.26 & 3 \\
Si I  & $ -0.35 $ & 0.26  & 5 & $ -0.28 $ & 0.09 & 5 \\
Ca I  & $ -0.49 $ & 0.12  & 5 & $ -0.47 $ & 0.21 & 4 \\
Sc II & $ -0.67 $ & 0.05  & 2 & $ -0.67 $ & 0.20 & 2 \\
Ti I  & $ -0.19 $ & 0.13  & 7 & $ -0.37 $ & 0.14 & 7 \\
Ti II & $ -0.31 $ & 0.07  & 2 & $ -0.46 $ & 0.06 & 2 \\
V  I  & $ -0.21 $ & 0.16  & 2 & $ -0.44 $ & 0.05 & 2 \\
Cr II & $ -0.32 $ &       & 1 & $ -0.35 $ &      & 1 \\ 
Mn I  & $ -0.41 $ &       & 1 & $ -0.30 $ &      & 1 \\
Fe I  & $ -0.28 $ & 0.16  & 15& $ -0.21 $ & 0.19 & 15 \\
Fe II & $ -0.33 $ & 0.14  & 3 & $ -0.23 $ & 0.04 & 4  \\
Co I  & $ -0.34 $ & 0.21  & 2 &  --       & --   & -- \\
Ni I  & $ -0.56 $ & 0.16  & 6 & $ -0.44 $ & 0.21 & 6  \\
Cu I  & $ -0.45 $ &       & 1 & $-0.22  $ &      & 1  \\ 
Y II  & $ -0.61 $ & 0.08  & 3 & $-0.66  $ & 0.13 & 4  \\
Ba II & $ -0.08 $ &       & 1 & $-0.10  $ &      & 1  \\
La II & $ +0.30 $ & 0.30  & 3 & $+0.33  $ & 0.23 & 3  \\
Ce II & $ +0.15 $ & 0.06  & 3 & $+0.00  $ & 0.13 & 3  \\
Nd II & $ +0.12 $ & 0.27  & 5 & $+0.09  $ & 0.19 & 8  \\
Eu II & $ +0.05 $ &       & 1 & $+0.05  $ &      & 1  \\
\hline
\end{tabular}

\end{center}
\end{table}

\begin{table}[]
\caption{Line data and abundances for Fe}
\begin{center}
\renewcommand{\tabcolsep}{0.1cm}
\begin{tabular}{lrrrrrr}
\hline
Ion   & $\lambda $ & log gf & EW(pm) & $\epsilon$ & EW(pm) & $\epsilon$ \\  
      &    nm     &        &  139&               &  143 &   \\
\hline
Fe I  & 585.5091  &  -1.76  & 4.12   & 7.56      &  4.24   & 7.61 \\
Fe I  & 585.6083  &  -1.64  & 3.85   & 7.04      &  6.24   & 7.51 \\
Fe I  & 585.8779  &  -2.26  & 1.97   & 7.14      &  2.37   & 7.27 \\
Fe I  & 586.1107  &  -2.45  & 1.81   & 7.35      &  1.22   & 7.18 \\
Fe I  & 506.7151  &  -0.97  & 7.46   & 7.06      &  8.34   & 7.24  \\
Fe I  & 510.4436  &  -1.69  & 5.89   & 7.27      &  6.58   & 7.65 \\
Fe I  & 510.9650  &  -0.98  & --     & --        &  7.90   & 7.24 \\
Fe I  & 489.2871  &  -1.29  & 7.09   & 7.32      &  5.91   & 7.06 \\
Fe I  & 552.5539  &  -1.33  & 6.25   & 7.13      &   -     &  --  \\
Fe I  & 587.7794  &  -2.23  & 2.74   & 7.25      &  3.34   & 7.25  \\
Fe I  & 588.3813  &  -1.36  & 7.88   & 7.18      &  9.15   & 7.43  \\
Fe I  & 615.1617  &  -3.30  & 9.59   & 7.28      &  8.02   & 6.98  \\
Fe I  & 616.5361  &  -1.55  & 5.98   & 7.17      &  6.46   & 7.26  \\
Fe I  & 618.7987  &  -1.72  & 7.36   & 7.37      &  6.34   & 7.17  \\
Fe I  & 649.6469  &  -0.57  & 6.26   & 6.97      &  8.45   & 7.38  \\
Fe I  & 670.3568  &  -3.16  & 7.72   & 7.42      &  7.16   & 7.33  \\
Fe II & 483.3197  &  -4.78  & 2.72   & 7.34      &  1.94   & 7.27  \\
Fe II & 499.3358  &  -3.65  & 5.97   & 7.10      &  6.37   & 7.30  \\
Fe II & 513.2669  &  -4.18  &  --    & --        & 3.98    & 7.33 \\
Fe II & 526.4812  &  -3.19  & 6.00   & 7.11      &  6.09   & 7.24  \\  

\hline
\end{tabular}
\end{center}
\end{table}

\section{Observations and data reduction}

\begin{figure*}
\psfig{figure=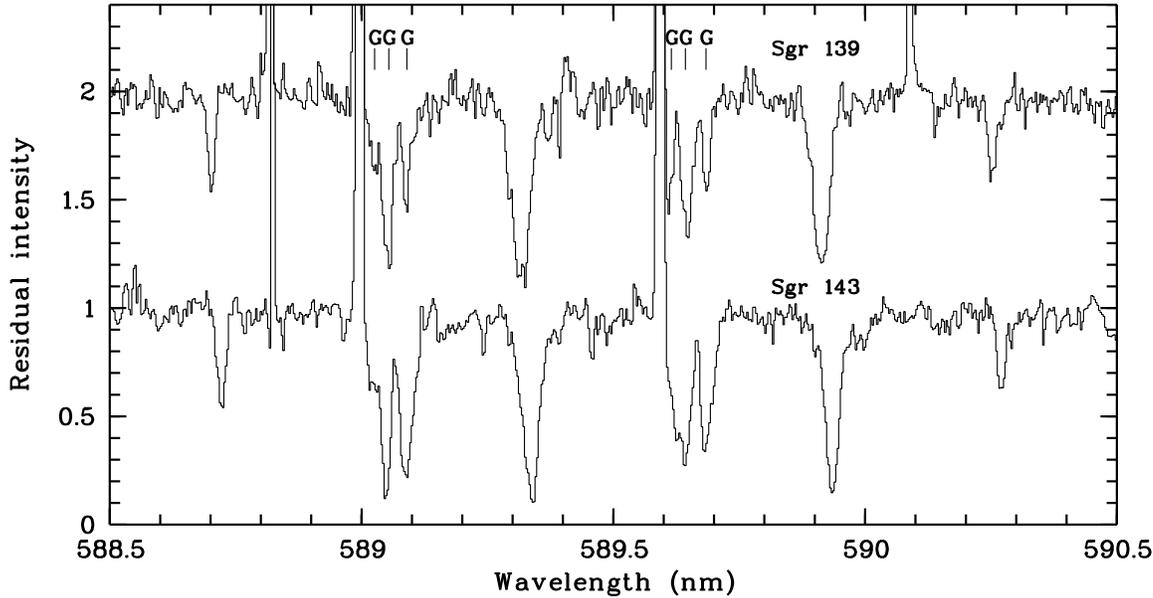,width=15.2cm,clip=t}
\caption{The spectra of the two stars
in the region of the Na I D doublet. The spectrum
of Sgr 139 has been shifted vertically by one unit for display
purposes. Wavelengths are observed wavelengths, not shifted to
rest. The absorptions due to Galactic interstellar gas
are clearly visible and are marked with the label G.}
\label{na}
\end{figure*}

The data were obtained during the Commissioning of UVES and have been
released by ESO for public use.
The spectra 
of the two stars, whose basic data
is given in Table 1,
were taken on the nights 2,3,4 and  6 October 1999.
The slit was $1 ''$, which 
provided a resolution of $\sim 43000$ at 565 nm.
We used only the red arm of UVES with the standard setting
with central wavelength at 580 nm,
which provides spectral coverage from  480 nm  to 680 nm  
The detector was the  mosaic of two CCDs
composed of a EEV CCD-44 for the
bluemost part of the echellogramme and
an MIT CCID-20 for the redmost part.
Both CCDs are composed of $4096\times 2048$ square pixels of 15 $\mu m $ side.
We used a $2\times 2 $ on--chip binning, without any loss of resolution,
given the relatively wide slit used. 
For each of the two stars three one-hour exposures were collected,
under
median seeing conditions, allowing to reach a signal to noise ratio of
$\sim 30$ at 510 nm, confirming the excellent performance
of the instrument (D'Odorico et al, 2000).

The data was reduced using the {\tt ECHELLE} context of {\tt MIDAS};
reduction included background subtraction, cosmic ray filtering,
flat fielding, extraction, wavelength calibration and order merging.
Each of the 
CCDs of the mosaic was reduced independently. 
The r.m.s. of the calibration was typically of the order of 0.2 pm for
each order and in all cases less than 0.3 pm.
Flat-fielding was highly successful and the single echelle orders
were rectified to within $\sim 5$\% by this process, except in the vicinity
of CCD defects, where the correction was not always satisfactory.
The three spectra available for each star were coadded without
any shift in wavelength. By cross-correlation we  estimated
the shift between
any pair of spectra to be less than 0.5 pixel, so we decided not to
perform any shift. This results in a very slight degradation 
of the resolution, which is not an issue for our analysis.
The differences in barycentric correction were at most of 0.1 kms$^{-1}$,
so no appreciable shift was expected from this cause either.
The normalized and merged spectra 
were plotted superimposed on preliminary
synthetic spectra for the purpose of line
identification.
A single velocity shift was adequate
for all the spectra, confirming that the internal
accuracy of our wavelength scale
is  better than 0.2 km/s. 
Portions of the normalized spectra of the two stars are 
displayed in figures 1 and 2. 

\section{Analysis}

Our analysis is standard and essentially based on LTE model atmospheres.
For each star we estimated effective temperature from 
$(V-I)_0$ 
through the colour--T$\rm _{eff}$ calibration for
giants of Alonso et al (1999).
We  adopted log g = 2.5 for both 
stars, this  value is 
compatible with the position of the stars
in the colour--magnitude diagram,
whichever the isochrone set considered.
Furthermore we verified {\it a posteriori} that
this gravity nearly satisfied the ionization equilibrium of both
Fe I/Fe II and Ti I/Ti II . 
The atmospheric parameters are summarized in Table 1.
The mean abundances for all
the elements are given in Table 2. 

With these parameters we computed model atmospheres using the ATLAS9
code  (Kurucz 1993) using opacity distribution functions with
microturbulence of 2 km/s and suitable metallicity.

We began by deriving the Fe abundance for both stars. We picked a set of
lines which our preliminary synthetic spectra predicted to be substantially
unblended, which had accurate laboratory or theoretical gf values
and which spanned a range of line strengths.
We excluded from the analysis lines stronger than $\log (EW/\lambda)=-4.8$
to avoid
an excessive dependence on microturbulence and lines weaker than
$\log (EW/\lambda)=-6.1$ to avoid lines which are too noisy.

For these lines we measured equivalent widths by fitting a gaussian with
the {\tt iraf} task {\tt splot} and used these and the model atmosphere
as input to the WIDTH9 code (Kurucz 1993). 
Some of the lines were removed from
the analysis because they provided highly discrepant abundances. 
The microturbulence was determined by imposing that strong lines
and weak lines give the same abundance.
The results are given in Table 3. 
The excitation equilibrium for Fe I is very nearly satisfied
for both stars (slopes are -0.05 dex/eV 
for star 139 0.13 dex/eV for star 143), we did not  adopt
an excitation temperature, so that the equilibria are better satisfied,
since our lines cover a range of only about 2.7 eV, this means, that in the
worst case (star 143) the slope predicts a difference of slightly over
0.3 dex between the highest and lowest excitation lines, this is 
$\sim 2 \times $r.m.s.

As a check of our metallicity we derived Fe abundances also 
using MARCS models computed by Plez et al (1992).
The difference is not significative: for both stars
the MARCS models provide an Fe I abundance which is 0.02 dex
larger than that provided by the ATLAS models, while for
Fe II it is 0.07 dex larger for star 139 and 0.05 dex larger
for star 143. 

Having fixed the metallicity and the microturbulence we proceeded
to determine all the other abundances with  the same method,  
again we disregarded the too strong and too weak lines, with
a few exceptions, such as  Sc, Cu and Ba, for which only one
or two lines were available,  in order
to get information on as many elements as possible.
The line data and abundances are given in Table 4.
In addition for some blended lines we resorted to spectrum synthesis
using the same model-atmosphere and the  SYNTHE code (Kurucz 1993).
We did not take into account hyperfine splitting (HFS)
for Sc, V, Mn, Co and Cu, however given that the abundances of
these elements
are coherent with those of other elements we do not 
expect corrections due to HFS to be very large.
For Eu we determined the abundance from the Eu II 
664.5 nm line, taking into account HFS splitting, the relevant data
is given in Table 5.
The error on the Eu abundances estimated from the quality of the fit is 
0.15 dex.

The main result of this analysis confirms the impression gathered by
a direct comparison of the spectra of the two stars: the stars are very nearly
identical, the few differences in their spectra are quite likely
determined by slightly different T$\rm _{eff}$ and log g, 
but not by chemical composition.

From the measure of the line centers 
of the unblendend lines used for abundances we
determined the radial velocity for the two stars. 
We obtain 
the following heliocentric radial velocities
$133.8 \pm 0.8 $ kms$^{-1}$ from 60 lines
for star 139 and $143.8 \pm 0.8 $ kms$^{-1}$
from 57 lines for star 143, the quoted error is just the r.m.s.
The measurement of the position of the atmospheric Na I D emission lines
allowed to estimate the zero point shift to be less than 0.1  kms$^{-1}$,
this, coupled with the excellent reproducibility of the wavelength
scale from night to night induced us to assume a null zero--point shift.
These heliocentric radial velocities support membership to Sagittarius, 
Ibata et al (1995) give a mean heliocentric
radial velocity for Sgr of $140 \pm 2 $  kms$^{-1}$
and Ibata et al (1997) find the intrinsic velocity dispersion
to be $11.4 \pm 0.7$  kms$^{-1}$ and
constant across the face of the galaxy. 
The N--body model of Sgr computed by Helmi \& White (2000)
displays a similar  velocity dispersion, if only the stars
with 100 kms$^{-1}\le v_{hel} \le 180$ kms$^{-1}$ are included,
as done by Ibata et al. However if this condition is relaxed
the velocity dispersion turns out to be much larger, due
to the contribution of stars in the debris streams.
Given the above considerations it is not surprising that we find
a difference of 10 kms$^{-1}$ between our stars.
Our measured  radial velocities
compare quite well with those measured from 
our EMMI low resolution spectra (147  kms$^{-1}$ for star 139 and
154  kms$^{-1}$ for star 143, both are accurate to $\pm 15 $ kms$^{-1}$).

We cannot rule out the possibility that the stars studied here
belong to the Bulge, rather than Sgr. If they were at a distance of
8.5 Kpc, rather than 25 Kpc their log g should be $\sim 0.5 $ dex
higher than what we assumed, but such a difference is within
the errors of the analysis. However the radial velocity ought to be
a very good discriminant. By looking at Figure 1 of Ibata et al (1995)
we see that the distribution of radial velocity of Bulge stars in
directions which do not intercept the Sgr dSph, show a vanishingly
small number of stars at the radial velocity of Sgr.
Also the chemical composition suggests that the two stars do not
belong to the Bulge: in fact Bulge stars are expected, theoretically,
to have [$\alpha$/Fe]$>0$, even at solar metallicities. Observationally
the situation is not so clear, however our distinctly solar [$\alpha$/Fe]
is a clue against Bulge membership.

\section{Discussion.}

 The metallicity of the two stars
examined here is higher than all previous 
photometric estimates. 
Although it is possible that we happened to select
two members of the high--metallicity tail of Sgr, this
position is hardly tenable, 
the event of finding two such stars
in a 9 square arcmin field must be quite rare.
It is more likely that Sgr actually possesses a population,
perhaps the main population, this metal--rich.
For our two stars the Schlegel et al (1998) maps
   provide $E(B-V)= 0.14$.
By comparison Marconi et al (1998)
used E(B-V)=0.18. 
The fact that the actual reddening could be 0.04 less than this could explain
why the metallicity we find is 0.3 dex higher than the highest metallicity
estimated by Marconi et al.

\begin{table}[]
\caption{Line data and abundances}
\begin{center}
\renewcommand{\tabcolsep}{0.1cm}
\begin{tabular}{lrrrrrr}
\hline
Ion   & $\lambda $ & log gf & EW(pm) & $\epsilon$ & EW(pm) & $\epsilon$ \\  
      &    nm     &        &  139&               &  143 &   \\
\hline
O  I  &  630.0304 & -9.82 & $<1.5 $ & $<8.33$    & 2.08    &  8.49   \\    
Na I  &  498.2814 & -0.95 & --      &  --        & syn      &  5.44   \\
Na I  &  615.4226 & -1.56 & 3.73    &  5.80      &  3.32    &  5.74   \\
Na I  &  616.0747 & -1.26 & 4.48    &  5.63      &  3.42    &  5.46   \\
Mg I  &  571.1088 & -1.83 & 11.48   &  7.32      &  10.16   &  7.12   \\
Mg I  &  631.8717 & -1.98 & 4.00    &  7.17      &  3.19    &  7.03   \\
Mg I  &  631.9237 & -2.20 &  --     &  --        &  3.24    &  7.26   \\
Al I  &  669.8673 & -1.65 & 2.88    &  5.95      &   syn    &  6.27   \\
Al I  &  669.6788 & -1.42 & syn     &  6.04      &  syn     &  6.17   \\
Al I  &  669.6023 & -1.35 & syn     &  5.87      &  syn     &  5.77   \\
Si I  &  577.2146 & -1.75 & 6.67    &  7.48      &  4.82    &  7.16   \\
Si I  &  594.8541 & -1.23 & 7.12    &  7.00      &  8.84    &  7.28   \\
Si I  &  614.2483 & -1.48 & 2.36    &  7.05      &  4.34    &  7.40   \\
Si I  &  614.5016 & -1.43 & 4.99    &  7.47      &  3.79    &  7.24   \\
Si I  &  615.5134 & -0.77 & 5.80    &  6.95      &  7.87    &  7.27   \\
Ca I  &  615.6023 & -2.20 & 1.81    &  5.77      &  --      &  --     \\
Ca I  &  616.1297 & -1.02 & 7.99    &  5.89      & 9.20     &  6.00   \\
Ca I  &  616.6439 & -0.90 &  --     &   --       & 8.01     &  5.66   \\
Ca I  &  645.5598 & -1.35 & 6.37    &  5.81      & 7.85     &  6.07   \\ 
Ca I  &  649.9650 & -0.59 & 9.95    &  5.70      & 9.95     &  5.69   \\  
Ca I  &  650.8850 & -2.11 & 3.28    &  6.01      & ---      &  --     \\
Sc II &  552.6790 & 0.13  & 9.94    &  2.53      &  10.91   &  2.65      \\
Sc II &  632.0851 & -1.77 & 1.92    &  2.46      & 1.62     &  2.36  \\
Ti I  &  488.5082 & 0.36  & 9.92    &  4.88      & 8.97     &  4.63  \\
Ti I  &  497.7719 & -0.92 & syn     &  4.66      & syn      &  4.80  \\
Ti I  &  497.8222 & -0.39 & syn     &  4.76      & syn      &  4.60  \\
Ti I  &  498.9131 & -0.22 & syn     &  4.76      & syn      &  4.44  \\
Ti I  &  499.7098 & -2.12 & syn     &  5.05      & syn      &  4.70  \\
Ti I  &  586.6452 & -0.84 &  10.14  &  4.91      & 8.41     &  4.57  \\
Ti I  &  612.6217 & -1.42 &  6.40   &  4.76      & 6.72     &  4.83  \\
Ti II &  498.1355 & -3.16 &  syn    &  4.76      & syn      &  4.60  \\  
Ti II &  499.6367 & -2.92 &  syn    &  4.66      & syn      &  4.52  \\
V  II &  573.7059 & -0.74 &   --    &  --        &  3.28    &  3.57  \\
V  II &  613.5361 & -0.75 &  syn    &  3.68       &  syn    &  3.50  \\
V  II &  615.0157 & -1.78 &   syn   &  3.91       &  syn    &  3.60  \\
Cr II &  488.4607 & -2.08 & 3.76    &  5.35       &  3.76   &  5.32  \\
Mn  I &  511.7934 & -1.14 & 3.72    &  4.98       &  4.24   &  5.09  \\
Co  I &  553.0774 & -2.06 & 5.76    &  4.73       &  --     &  --    \\
Co  I &  533.1452 & -1.96 & 4.10    &  4.43       &  --     &        \\
Ni  I &  585.7746 & -0.64 & 2.74    &  5.41       & 4.38    &  5.75  \\
Ni  I &  612.8963 & -3.33 & 5.60    &  5.75       & 5.86    &  5.81  \\
Ni  I &  613.0130 & -0.96 & 1.89    &  5.61       & 1.88    &  5.62  \\ 
Ni  I &  617.5360 & -0.53 & 5.74    &  5.81       & 4.93    &  5.64  \\
Ni  I &  617.6807 & -0.53 & 5.73    &  5.81       & 7.86    &  6.21  \\
Ni  I &  617.7236 & -3.50 & 3.69    &  5.75       & 3.95    &  5.82  \\
Cu  I &  510.5537 & -1.52 & 13.10   &  3.99       & 11.80   &  3.76  \\
Y  II &  488.3684 & 0.07  & 5.84    &  0.95       & 9.37    &  1.67  \\
Y  II &  498.2129 & -1.29 & 3.35    &  1.71       & 2.78    &  1.57  \\
Y  II &  508.7416 & -0.17 & 7.58    &  1.55       & 7.13    &  1.40  \\
Y  II &  511.9112 & -1.36 & 2.91    &  1.63       & 3.15    &  1.67  \\
Ba II &  649.6897 & -0.38 & 18.80   &  2.06       & 18.92   &  2.03  \\
La II &  480.4039 & -1.50 & 3.76    &  1.40       & 3.53    &  1.33  \\
La II & 511.4559  & -1.06 & 7.74    &  1.81       & 7.78    &  1.76  \\
La II & 632.0376  & -1.61 & 3.20    &  1.21       & 4.31    &  1.42  \\
Ce II & 518.7458  &  0.13 & 4.30    &  1.74       & 3.00    &  1.44  \\
Ce II & 533.0556  & -0.36 & 3.53    &  1.66       & 3.26    &  1.59  \\
Ce II & 546.8371  &  0.14 & 3.60    &  1.78       & --      &  --    \\
Ce II & 604.3373  & -0.43 &  --     &  --         & 2.04    &  1.71  \\
Nd II & 491.4382  & -1.00 & 5.89    &  2.01       & 3.82    &  1.53  \\
Nd II & 495.9119  & -0.98 & 11.66   &  $^n$2.87       & 6.73    &  1.75  \\
Nd II & 496.1387  & -0.71 & 2.80    &  1.32       & 3.01    &  1.36  \\
Nd II & 498.9950  & -0.50 &  --     & --         & syn      &  1.79  \\
Nd II & 499.8541  & -1.10 &  --     &  --        & syn      &  1.79  \\
Nd II & 508.9832  & -1.16 & 3.79    & 1.49       & 2.94     &  1.30  \\
Nd II & 529.3163  & -0.10 & 5.70    & 1.54       & 6.48     &  1.67  \\
Nd II & 543.1516  & -0.57 &  --     & --         & 2.84     &  1.71  \\
Nd II & 548.5696  & -0.30 & 3.53    & 1.77       & 2.32     &  1.48  \\

\hline
\end{tabular}

\noindent{ $^n$   not  used to compute the mean abundance}
\end{center}
\end{table}

\begin{table}[]
\caption{HFS data for Eu II}
\begin{center}
\begin{tabular}{lrr}
\hline
 & $\lambda$ (nm) & log gf \\
\hline
 Eu II & 664.516 &$  -0.800$\\
 Eu II & 664.513 &$  -0.886$ \\
 Eu II & 664.511 &$  -0.521$ \\
 Eu II & 664.510 &$  -0.856$ \\
 Eu II & 664.509 &$  -0.506$ \\
 Eu II & 664.508 &$  -0.771$ \\
 Eu II & 664.507 &$  -0.427$ \\
\hline
\end{tabular}

\end{center}
\end{table}

Quite obviously our results do not  
rule out the existence of a more metal--poor population.
Preliminary results of abundance analysis in Sgr are given 
by Smecker-Hane \& McWilliam (1999),  who 
find two metal--poor  Sgr member stars,
with [Fe/H]$=-1.41$ and [Fe/H]$=-1.14$.
It is interesting that out of 11 stars analyzed by them
7 have metallicities in the range $-0.6 < \rm [Fe/H] < -0.2$,
two are metal-poor and two are metal rich ([Fe/H]$\sim 0.$).
The 7 stars of intermediate metallicity, which should be analogous
to the two under study here, show solar abundance ratios
and no enhancement of $\alpha$ elements, in agreement with our
findings. 
Also the Na abundance displays a similar pattern: for all their stars,
except the two metal--poor ones  Na is over-deficient with
respect to iron by 0.3 -- 0.5 dex. 
Unfortunately, these results have not
been published in a more detailed form and we lack information
on the temperatures and luminosities
of 
the the stars considered by Smecker-Hane \& McWilliam
so we do not know if we are comparing similar giants.

It is also  interesting to compare the present results with
the abundances    of the two Sgr planetary nebulae
He 2-436 and Wray 16-423, studied by Walsh et al (1998).
The only element in common in the two analysis is O, for which
Walsh et al find [O/H]$= -0.64 \pm 0.08$ and [O/H]$=-0.62 \pm 0.07$
for He 2-436 and Wray 16-423, respectively.
Our result for Sgr 143 
is about 0.2 dex higher, but it is also more uncertain, because it
is based on a single weak line, which is also very sensitive to gravity.
An increase of gravity of 0.5 dex results in an increase
of O abundance of 0.25 dex.
O should be only marginally affected during AGB evolution,
so that the O abundance in the PN ought to be quite close to
that in the progenitor star. Walsh et al (1998) argued
that their abundances suggested a mild enhancement of O over Fe,
because they assumed $-0.8$ to be the mean [Fe/H] of Sgr.
Another scenario appears more likely, in view of 
our results: a solar O/Fe ratio, which suggests
that the PNe have [Fe/H]$\sim -0.5$.

Having established that the two stars are quite similar in
atmospheric parameters and abundances we must explain why their photometry 
is different and why the metallicity estimated from the low
resolution spectra for star 143 is far lower than the 
one derived here.
We consider 5 possibilities: 1) errors in $V$;
2) errors in $V-I$; 3) different reddening; 4) different age;
5) different distance. Let us  examine all of these cases.

That a difference of 0.18 mag in $V$ may be due to the
photometric error may be discarded since this is a factor of ten
larger than the photometric error of Marconi et al (1998).
An error in $V-I$ is more likely;  a 0.03 -- 0.04 mag error in $V-I$
would allow 
to slide sideways one of the two stars in the colour-magnitude diagram
in such a way that both stars lie on the same isochrone, since the RGB, in
this range of $V-I$ is very steep. The implied difference in
T$_{\rm eff}$ is of $\sim 100$ K,  the errors of our analysis.

Differential reddening seems unlikely for three reasons.
The dust maps of Schlegel et al (1998) give $E(B-V)=0.14$
   for both stars,
suggesting that the reddening of the two stars is the same within 0.01 mag.
The absence of detectable amounts of HI in Sgr (Burton \& Lockman, 2000)
also argues
against a differential reddening.
If the 0.18 mag difference in $V$ were due to reddening
it would imply a difference of almost 0.08 mag
in $V-I$, i.e. $\sim 200$ K in T$\rm _{eff}$.
Although such a difference is within the errors of the present
analysis and cannot be ruled out, it does seem somewhat
unlikely, given the similarity of the two spectra.

An age difference of $\sim 1 $ Gyr could be enough to explain the
difference in the photometry of the two stars. A larger age spread
would be necessary to explain the width of the RGB, like in the scenario
proposed by Bellazzini et al (1999). Although such a
possibility is attractive, it appears somewhat contrived and it is not
so clear that star formation may continue for several Gyrs without
resulting in a spread in metallicity, as well as ages.

A distance difference
of about 2Kpc would be enough to explain the difference
in $V$.
This value is  not unreasonable,
Ibata et al (1997), estimate the half-brightness depth of Sgr to be about
1.2 kpc . 
It is interesting that recent N--body simulations by Helmi \& White
(2000) support a considerable depth of Sgr:
inspection of their figure 2
shows that the bulk of their model for
Sgr has a depth of about 2 Kpc, however
considering the debris shed during previous orbits, one has a sizeable
population over a depth of 10 Kpc.
Further support to the possibility 
that the two stars have a different distance
comes from inspection of the Na I D lines (Fig. 2),
three interstellar components belonging to our Galaxy are
evident in both the spectra of Sgr 143 and of Sgr 139
at radial velocity +16.5 kms$^{-1}$, $+28.0$  kms$^{-1}$
and $+47.3$ kms$^{-1}$;
while the Na I D lines of star 143 appear symmetric and there
is no hint of an interstellar component at the radial
velocity of Sgr, the lines of star
139 show a weak but definite asymmetry, which we interpret
as a weak interstellar line associated with Sgr.
Star 139 is in fact the fainter of the two and hence the most
distant, according to this interpretation,  this would explain
why the interstellar Na I D lines appear in its spectrum but
not in the spectrum of star 143, which would then be in the side
of Sgr nearer to us.

So of the five possibilities considered only the photometric
error in $V$ and the differential reddening are discarded.
We may not decide which is the correct one with the present
data, new accurate photometric measurements will allow
to settle at least the issue of errors in $V-I$. 
However we consider that the  distance difference
is the most likely explanation, because it is the simplest
and is supported by several arguments. This
suggests that the non-negligible line of sight depth of Sgr
could explain at least a part of the width of the RGB
of Sgr.
Up to now all investigators have adopted a unique
distance modulus for Sgr, in order to compare their photometry
to fiducial ridge lines of Galactic clusters or to theoretical
isochrones. This assumption may prove to be  bit too naive.

A full discussion of the 
metallicity estimates from low resolution spectra
shall be given elsewhere.
Suffice to say here that the method of estimating abundances
from low resolution needs 
a relatively high S/N ratio.
In the case of star 139
the metallicity derived from high resolution analysis coincides with that
estimated from low resolution to within the errors of the latter.
We verified that   
the degraded UVES spectrum is very similar to the low resolution EMMI 
spectrum. The indices measured on this degraded spectrum yield 
in fact almost the same abundance provided by those measured on the
low resolution spectrum. 
In the case of  star 143 instead   the method
has been applied to a spectrum of too low signal to noise ratio, in this
case the degraded UVES spectrum bears almost no resemblance to the
low resolution spectrum, 
except for the strongest  feature of the Mg I b triplet,
which was enough to provide the correct radial velocity
for this star.

The ratios of all elements are essentially solar,
noticeable exceptions are: Na which is overdeficient
with respect to iron, and the heaviest elements Ba, La, Ce, Nd, Eu,
which appear over-abundant while Y appears underabundant.
Such anomalies are not readily interpretable, deep mixing
would produce an enhanced Na and low O and Mg, at variance to what is
observed.
While it would be tempting to interpret
the overabundance of heavy elements as due to s-process
enrichment, the stars do not appear luminous enough to 
be on the thermally pulsating asymptotic giant
branch, where this mechanism is operative.
Moreover, s-process
enrichement would produce also a high Y abundance and no Eu 
(which is thought to be a ``pure'' r-process element), 
at variance to the low Y and high Eu abundances observed here. 
On the other hand, these stars could have been born in r-process
enhanced
material (sugested by the Eu enhancement). Howevever this seems also
quite
implausible since the r-process is thought to take place in Type II
supernovae
which also produce large amounts of O and other
$\alpha$-elements,  which are not observed to be enhanced in our stars.
This surprising pattern is reminiscent to what is observed in the
young supergiants in both
Magellanic Clouds, where the ratios of the moderate-mass s-process
elements Y and Zr
to iron are essentially solar, whereas the heavier species Ba to Eu
are overabundant by ratios [X/Fe] of the order of 0.3 and 0.5~dex 
respectively in the LMC and SMC (Hill et al. 1995, Hill 1997, Luck et
al 1998).  In the Magellanic Clouds also, we are at loss of an
explanation for this behaviour (see discussion in Hill 1997).
Note that our two Sgr giants have the same overall
metallicity as the LMC young population, and that the heavy elements
overabundances are also of the same order as in the LMC.

\begin{acknowledgements}
We are extremely grateful to 
S. D'Odorico, H. Dekker and the whole UVES team for  
the conception and construction of 
this wonderful
instrument.
\end{acknowledgements}

\end{document}